# On the upper limit of laser intensity attainable in non-ideal vacuum

Yitong Wu,[1,2,3] Liangliang Ji,[1,3,*] and Ruxin Li[1,3,4, †]

[1]*State Key Laboratory of High Field Laser Physics, Shanghai Institute of Optics and Fine Mechanics, Chinese Academy of Sciences, 201800 Shanghai, China*

[2]*Center of Materials Science and Optoelectronics Engineering, University of Chinese Academy of Sciences, 100049 Beijing, China*

[3]*CAS Center for Excellence in Ultra-intense Laser Science, 201800 Shanghai, China*

[4]*Shanghai Tech University, 201210 Shanghai, China*

[*] *jill@siom.ac.cn*

[†] *ruxinli@mail.siom.ac.cn*

**Abstract:** The upper limit of the laser field strength in perfect vacuum is usually considered as the Schwinger field, corresponding to ~$10^{29}$W/cm$^2$. We investigate such limitations under realistic non-ideal vacuum conditions and find out that intensity suppression appears starting from $10^{25}$W/cm$^2$, showing an upper threshold at $10^{26}$W/cm$^2$ level if the residual electron density in chamber surpasses $10^9$cm$^{-3}$. This is because the presence of residual electrons triggers the avalanche of quantum-electrodynamics cascade that creates copious electron and positron pairs. The leptons are further trapped within the driving laser field due to radiation-reaction, which significantly depletes the laser energy. The relationship between the attainable intensity and the vacuity is given according to particle-in-cell simulations and theoretical analysis. These results answer a critical problem on the achievable light intensity based on present vacuum conditions and provide a guideline for future 100's-Petawatt class laser development.

# 1. INTRODUCTION

Ultra-bright light sources have always been a major pursuit for their applications in various research areas. At the moment, femtosecond lasers based on the CPA (chirped pulse amplification) technique [1] is regarded as the most reliable approach to realize the highest peak power. After being amplified, compressed and focused, the peak laser intensity can reach up to $10^{22}$W/cm$^2$ [2]. The 10 PW-class laser facilities, such as ELI [3], Apollo [4], Vulcan [5] and SULF [6], aim at boosting the focused intensity by another 10 folds. Ambitious plans of 100 PW-class are proposed [7-9] worldwide, where the peak intensities of $10^{25}$W/cm$^2$ are anticipated. Furthermore, efforts have also been paid in exploring new mechanisms to generate Exawatt-Zettawatt lasers [10-12]. At such extreme light intensities, particle acceleration towards 10-100 GeV for leptons [13, 14] and 0.1-10GeV/nucleon for ions [15-17] are to be expected, which may further motivate applications in advanced X/gamma-ray sources [18, 19] and ion cancer therapy [20]. The new interaction regime featured by radiation-reaction and quantum-electrodynamics (QED) processes can be probed [21, 22] (e.g., radiation reaction [23-25], electron-positron generation [26-29], QED cascade [30-33], etc.). Nuclear physics [34-36] as well as lab-astrophysics [37-39] will also benefit a lot from these extreme laser sources.

While high power lasers are under fast development, a central question regarding the ultimate laser intensities researchers can build arises [40]. Basically, the upper limitations for laser intensity in ideal vacuum condition is considered as the Schwinger field $E_s= 2\pi m_e^2 c^3/e\hbar \sim 1.32 \times 10^{18}$V/m [41]. The QED theory predicts that at such field strength virtual particle-pairs could gain enough energy before annihilation and become real particles. In other words, energy of the laser photons is transferred to electron-positron pairs and $\gamma$ ray photons. Previous studies have shown that even a single pair produced in vacuum by a laser field can lead to rapidly depletion of laser energy [42], i.e., the maximum light intensity is strictly restricted below $10^{29}$W/cm$^{-3}$ in vacuum.

In reality, it is impossible to build a perfect vacuum environment for experiments. Typically, the vacuum electron density in a chamber suitable for PW-class lasers is about $10^{11}$cm$^{-3}$, provided by ordinary pumping technique (e.g., $10^{-3}$ Pa for SULF [6]). For laser power above 100PW, the chamber volume is enlarged by more than ten folds, posing great challenge to the pump. Another potential drawback is the existence of electrons extracted from optical components (focusing mirror, plasma mirror, etc.) by the passing laser fields. These residual electrons could serve as seeds to trigger the QED processes when the laser field surpasses certain threshold. Specifically, during the laser-electron interaction, nonlinear Compton Scattering [43] following $e+n\omega \rightarrow e+\gamma$ will occur, where electrons absorb multiple laser photons and emit high energy $\gamma$ photons. The radiated $\gamma$ photons further interact with the strong laser field, generating electron-positron pairs via the nonlinear Breit-Wheeler process ($\gamma+n\omega \rightarrow e^++e^-$) [44]. These two reaction channels build up positive feedbacks, i.e., the amount of the pairs and $\gamma$ photons will be avalanche-like amplified, known as the QED cascade [45, 46]. Previous studies pointed out that laser depletion is quite significant for near critical density target at peak intensities above $10^{24}$W/cm$^2$ [30-33]. Nevertheless, regarding the rather low electron density (about 7~8 orders smaller) in vacuum, the specific restriction on the attainable laser intensity is still unclear. This is a key question that needs to be answered for developing lasers beyond 100PW peak power.

In order to find out the upper limit of peak laser intensity under non-ideal vacuum conditions, we carried out particle-in-cell (PIC) simulations by including the QED models responsible for the two major reaction channels. The simulation results show that the attainable peak intensity does depend on the vacuity. At electron density about $10^9$cm$^{-3}$, notable energy drain emerges from $10^{25}$W/cm$^2$ and the laser intensity is restricted to the order of $10^{26}$W/cm$^2$. This threshold is well interpreted by our theoretical analysis.

## 2. SIMULATION SETUP

Our investigation is based on two-dimensional (2D) PIC simulations using the code VLPL (Virtual Laser Plasma Lab) [47]. It has implemented a QED-Monte-Carlo model accounting for nonlinear Compton scattering and Breit-Wheeler processes. In our simulations, laser propagates from left side of a moving simulation window along the $x$ direction. The window size is 40μm($x$)×80μm($y$) resolved by 4000×1000 cells. We set 2 macro-particles in each cell. The laser beam is linearly polarized along the $y$ axis, following a Gaussian profile $E_L=aw_0/w(x)\cos^2[\pi(t-t_f)/2\tau_0]\exp[-r^2/w^2(x)]$ focused at $x_f$=240μm with normalized peak amplitude $a=eE/m\omega c$, where $m$ is the mass of electron, $c$ is the velocity of light in vacuum and $\omega$ is the laser frequency. Here $r^2=y^2+z^2$, the laser wavelength is $\lambda$=800nm, beam width $w_0=3\lambda=2.4\mu m$, $w(x)=w_0\{[(x-x_f)^2+x_R^2]/x_R^2\}^{1/2}$, Rayleigh length $x_R=\pi w_0^2/\lambda$, focusing time $t_f=x_f/c$ and pulse duration $\tau_0=10\lambda/c$=26.7fs, respectively. The peak laser field amplitude $a$ is varied from 1500 to 10000 while the vacuum electron density is tuned between $10^{11}$ and $10^{15}$cm$^{-3}$. The simulation time step is $\Delta t$=0.008$T_0$.

Two challenges should be addressed while carrying out these simulations: (i) initialization of the low-density electrons and (ii) the memory cost for generated new particles ($\gamma$ photons and electron-positron pairs). It should be noted that at extremely low electron densities (e.g., $10^{11}$cm$^{-3}$), the average weight of electrons $w$ located in one cell is much less than 1, i.e., it is not physical to start the simulations with simple homogeneous initialization. Therefore, we take the following initialization strategy: first, the particle weight $w$ is calculated once the electron density is given; then, a [0, 1] uniform distributed random value $r_a$ is generated, by which the weight of the macro-particle is set to $w = \text{int}(w) + rank(w - r_a)$, where $rank$ is the step function with $rank(x\geq0) = 1$ while $rank(x<0)=0$; finally, If $w$=0 no macro-particles will be placed in the cell. To mitigate the memory issue in simulation clusters particle merging is turned on when the macro-particle number per cell surpasses 4 [47]. Moreover, modelling the QED cascade processes via the Monte-Carlo algorithm and initialization of low-density plasma induce stochastic features. To avoid contingency of the stochastic effects, 10 simulation examples with randomly distributed seeds are carried out at each set of parameters.

## 3. RESULTS AND DISCUSSIONS

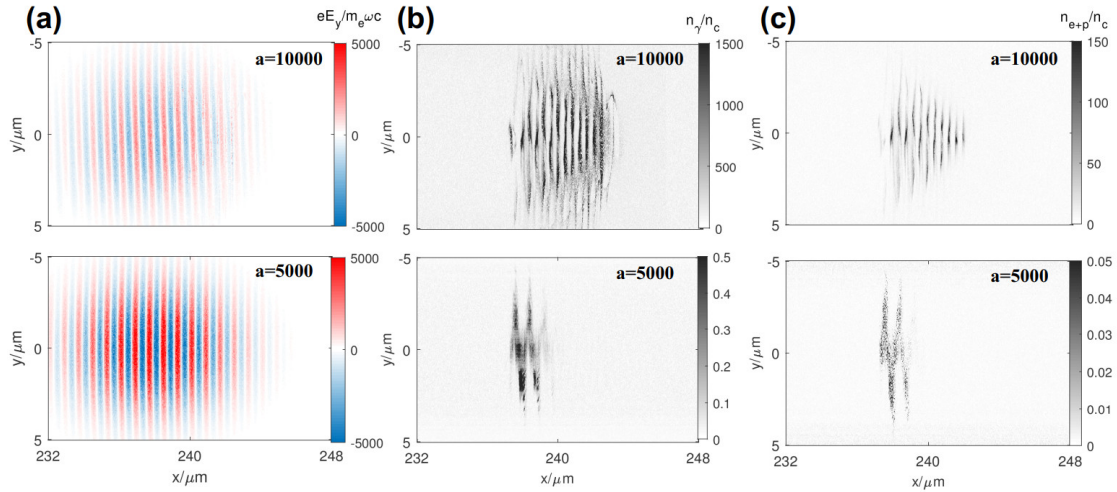

**Fig. 1.** The distributions of (a) laser electric fields $E_y$ (b) $\gamma$ photons density $n_\gamma$ as well as (c) electron-positron density $n_{e+p}$ at $t_f$=300$T_0$ and $n_{e0}$=$10^{11}$cm$^{-3}$ for $a$=10000 (top panel) and $a$=5000 case (bottom panel), respectively. The $E_y$ is normalized by $m_e\omega c/e$ while densities are normalized by critical density $n_c$.

We compare the results of $a$=10000 and 5000 at $n_{e0}$=$10^{11}$cm$^{-3}$ in Fig. 1. The peak laser field amplitude is well preserved for $a$=5000, as seen in Fig. 1(a). However, it declines to be less than 3000 for the other one. The remarkable difference indicates that the attainable light intensity at $n_{e0}$=$10^{11}$cm$^{-3}$ is subject to strong restrictions and the upper limit is appearing at $a$=10000. The density distributions of electron-positron pairs $n_{e+p}$ and $\gamma$ photons $n_\gamma$ are shown in Fig. 1(b) and 1(c), where both are about 3 orders of magnitudes higher for $a$=10000 case. The density profile shows distinctive patterns between the two cases. We notice that at $a$=10000 high density bunches appear all along the laser beam, while at $a$=5000 density peaks are only seen in the vicinity of highest laser intensity. This is because QED cascade is triggered at the rising edge of the laser pulse for the former such that copious electrons and positrons are created at an earlier moment.

When an ultra-intense laser interacts with background plasma, the ponderomotive force would usually expel local particles in both longitude and transverse directions. A channel is formed, where the laser pulse propagates through without significant volumetric energy drain. This is not the case in Fig. 1(b) and (c). We notice that electrons and positrons, mostly newly generated, sit within the intense laser beam. The corresponding phase-space density of electron-positron pairs are given in Fig. 2, by including the case of $a$=1000 for comparison. At relatively lower laser intensity ($a$=1000) transverse momenta dominate over the longitudinal one, meaning that a significant number of electrons are pushed away via laser ponderomotive scattering (see in Fig. 2(a)). The phase-space distribution is drastically different in the case of $a$=5000 and 10000. We see that the transverse momenta vanish for majority of the electrons/positrons, manifesting the clustering of leptons along the propagation axis, as displayed in Fig. 2(b) and (c). This phenomenon is known as radiation-reaction trapping (RRT) in travelling laser field [23], where the recoiling force of photon emission offsets the pondermotive force, leading to anomalous trapping of leptons in the most intense part of the laser field. It is consistent with the density distribution shown in Fig. 1(b). The threshold of RRT is around $a$=1000 according to previous studies [23-25]. In our case, the featured momentum distribution pattern appears when the laser amplitude goes beyond $a$=1000, in agreement with the theoretical predictions. Note that the RRT threshold is much smaller than the one required to seed QED cascade. This is particularly important in developing efficient cascading and laser energy depletion. If the thresholds are to be reversed, the generated particles would be expelled from the interaction region and the avalanche-like amplification would not sustain.

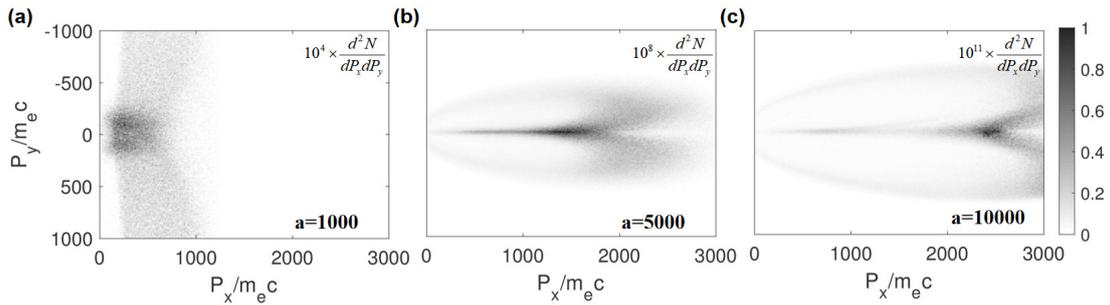

Fig. 2. The electron density in the momentum space $P_x$-$P_y$ at focusing time $t_f$ and $n_{e0}$=$10^{11}$cm$^{-3}$ for (a) $a$=1000, (b) $a$=5000 and (c) $a$=10000, respectively. The $P_x$, $P_y$ are normalized by $m_e c$.

In the following, we derive the theory that describes the evolution of particle numbers from QED cascade and give the criterion for laser energy depletion. We consider the $\gamma$ photon and electron-positron pair generation rates satisfying the expression:

$$\frac{dN_{e+p}}{dt} = 2\Gamma_e N_\gamma \quad (1)$$

$$\frac{dN_\gamma}{dt} = \Gamma_\gamma N_{e+p} - \Gamma_e N_\gamma \quad (2)$$

where $\Gamma_e$ and $N_{e+p}$ is the generated rate coefficient and number of electron-positron pairs; $\Gamma_\gamma$ and $N_\gamma$ are the coefficient and number of $\gamma$ photons, respectively. The generated rate of cascade processes are determined by the QED parameter $\chi_i=|(F_{uv}P_i^v)^2|^{1/2}/E_s m_e c$ [48] where $F_{uv}$ is the EM field tensor [49] and $P_i^v$ is the particle's four-momentum. According to previous research, the QED parameter can be approximated by $\chi_i \propto I^{3/4} \propto a^{3/2}$ [33, 50, 51] and generation rate $\Gamma$ is proportional to $\chi_i^{2/3}\exp(\chi_i^{-2/3}) \propto a\exp(-a)$ [51-53] for our simulations (Intensity $>10^{24}$W/cm$^2$). From empirical approximation that the $\Gamma_\gamma$ is about $1/T_0$ for $a_{ph}=0.01 a_s$ where $a_s$ is the normalized Schwinger field. Thus, we obtain the generated rates $\Gamma_\gamma \sim 4\Gamma_e$ [54] $\sim a\exp(-a_{ph}/a)/(a_{ph}T_0)$. Considering the energy of $\gamma$ photons and electron-positron pairs is about $am_e c^2/2$ [21, 53, 55], the laser depletion for such processes can be roughly evaluated as $dE \sim -ad(N_{e+p}+N_\gamma)m_e c^2/2$. Assuming the Gaussian profile remains the same during focusing $a=G(t)\xi=[1+(t-t_f)^2/t_R^2]^{-1/2}\xi$ and taking $dE \sim 2c_1 V_d \xi d\xi/c$ with $t_f=300T_0$, $t_R=x_R/c=9\pi T_0$, $V_d$ is the depletion region volume where $a>a_{RRT}$ and $c_1=m_e^2 c^3 \omega^2 \varepsilon_0/2e^2=1.38\times 10^{18}\times(1\mu m/\lambda)^2$ Wcm$^{-2}$, the evolution of $\xi$ is derived as follow:

$$\frac{d\xi}{dt} = -\frac{am_e c^3}{4c_1 V_d \xi}(4N_{e+p}+N_\gamma)\Gamma_e \quad (3)$$

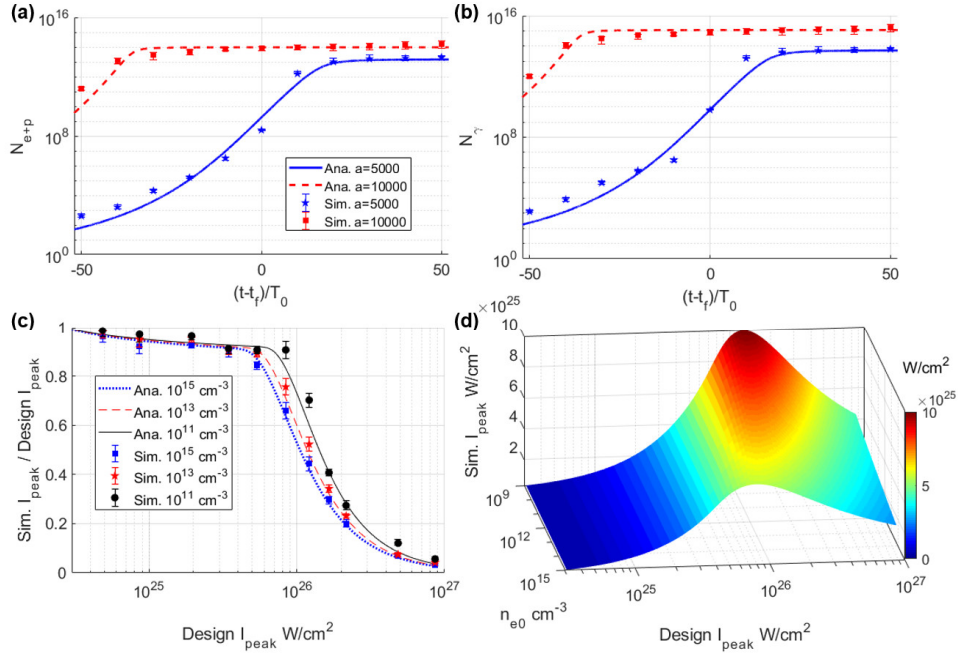

**Fig. 3.** (a) $N_{e+p}$ and (b) $N_\gamma$ evolution for $a=5000$ (blue solid and pentagons) and $a=10000$ (red dashed and circles) obtained from simulation (symbols) and theoretical analysis (lines). (c) The ratio between the measured peak intensity in simulations and the designed one as a function of designed peak intensity under electron densities of $n_{e0}=10^{15}$ (blue dotted and squares),$10^{13}$ (red dashed and pentagons),$10^{11}$cm$^{-3}$ (black solid and circles). The symbols are results measured from simulation while lines are from the theoretical model. All symbols represent average values for 10 simulation cases with different random seeds while the error bars represent peak intensity quantile of 95% and 5% (error bar gives confidence interval of 90%), separately. (d) The theoretical prediction of peak intensity distributions as a function of the designed peak intensity and $n_{e0}$ (from $10^9 \sim 10^{15}$cm$^{-3}$).

Combining Eqs. (1)~(3) and taking the cascade duration $t_a \sim [(a_0/a_{RRT})^2-1]^{1/2}t_R$ (corresponding to the period where $a_0>a_{RRT}$) with the initial conditions $N_{e+p}(t=t-t_f-t_a/2)=c\tau_0\pi w_0^2 n_{e0}$, $N_\gamma(t=t-t_f-t_a/2)=0$, the numerical solution of $N_{e+p}$, $N_\gamma$ and $a$ can be acquired. The evolution of $N_{e+p}$ and $N_\gamma$ based on the above analytical model are given in Fig. 3(a) and 3(b), together with the results collected from PIC simulations. The number of both electron-positron pairs and gamma photons undergo exponential growth when the laser interacts with residual electrons, owing to the avalanche-like cascade. When sufficient laser energy is drained, the light intensity declines (see the following discussion in Fig. 4) and the number of created particles saturates. The above trends are reproduced by our theoretical model.

The peak intensity during focusing processes are measured from PIC simulations and compared to our analytical model. Again, the results in Fig. 3(c) illustrate the consistency between the two. According to the systematic scanning, the reduction of peak intensity emerges from $10^{25}$W/cm$^2$, indicating that the depletion effects should be taken into consideration for above 100s' PW class laser facility. The ratio between the simulated peak intensity and the designed intensity decreases sharply when approaching $10^{26}$W/cm$^2$ for density from $10^{11}$ to $10^{15}$ cm$^{-3}$, corresponding to the energy depletion threshold. As seen in Fig. 3(d), when designed light intensity surpasses the threshold, the attainable one is restricted to $10^{26}$W/cm$^2$, exhibiting a clear ceiling. Our theoretical model indicates that the sharp boundary remains for vacuity down to even $10^9$cm$^{-3}$.

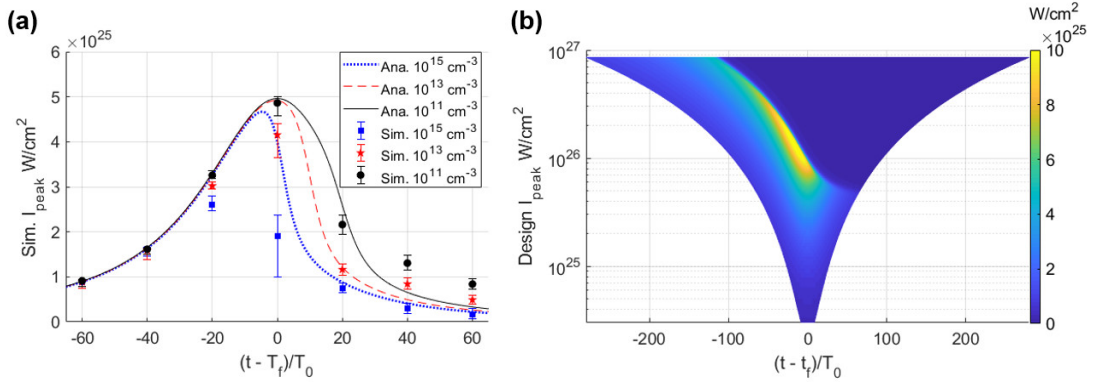

**Fig. 4.** (a) The obtained peak intensity evolution at $a$=5000 for different vacuum electron densities $n_{e0}$. The black circles, red pentagons and blue squares represent average peak intensity measured in simulations. The solid black line, dash red line and dotted blue line denote theoretical analysis with $n_{e0}=10^{11}, 10^{13}, 10^{15}$cm$^{-3}$, respectively. (b) The theoretical predicted peak intensity evolution from analytical model as a function of designed peak intensity at $n_{e0}=10^9$cm$^{-3}$.

In fact, the rising and falling edge of the laser pulse should be symmetric around $t=t_f$ in the time domain if depletion is negligible. Nevertheless, strong depletion breaks down the symmetrical profile such that the maximum intensity observed in simulations is not exactly at the designed focal position. The behavior is even more obvious with higher residual density (Fig. 4(a)) or at intensities beyond the threshold (Fig. 4(b)). We choose $a$=5000 as an example and present the peak intensity at different simulation time. As depicted in Fig. 4(a) from both simulations and the theoretical model, the laser peak intensity appears near the focal position for $10^{11}$cm$^{-3}$ while much earlier for the $10^{13}$cm$^{-3}$ (by ~$10T_0$) and $10^{15}$cm$^{-3}$ (by ~$25T_0$) cases. Since laser intensity at the pulse rising front does not reach the threshold at $10^{11}$cm$^{-3}$, the distortion caused by depletion is negligible. At higher electron densities, the intensity exceeds the threshold before $t=t_f$ and the cascade develops quickly. Significant depletion in the laser front induces intensity peak shifting to an earlier time than designed. The highest intensity found from simulations as a function of the propagation time is presented from our theoretical model for $n_{e0}=10^9$cm$^{-3}$. As one notices in Fig. 4(b), the symmetrical time profile of peak intensity becomes asymmetric when approaching the threshold. In this case, the attainable intensity is restricted to below $10^{26}$W/cm$^2$.

## 4. CONCLUSION

In summary, we have explored the attainable highest laser intensity under different vacuum conditions for the first time, to the best of our knowledge. It is found that the avalanche-like QED cascade and radiation-reaction trapping effect pose a strong limit on the achievable light intensity due to the residual electrons the laser pulses meet. Our study suggests that the observed peak intensity is suppressed starting from ~$10^{25}$W/cm$^2$ and an upper limit emerges at $10^{26}$W/cm$^2$ for vacuum electron densities above $10^9$cm$^{-3}$. These laser intensity thresholds can be approached by focusing the optical laser pulses of multiple hundreds of PW peak-power. The cases for building lasers beyond 100s' PW peak power are therefore not well justified when the required vacuum conditions are inaccessible by state-of-art techniques.

It is worth noticing that light intensities at ~$10^{24}$W/cm$^2$ can readily support the research on strong-field QED physics (e.g., the radiation-reaction effects, electron-positron pair production, QED cascade etc), particle acceleration towards the high-energy frontier, laser-driven nuclear physics and high-energy density physics. The featured intensity is already accessible with a 100-PW laser, such as the SEL 100PW laser under construction in China [56].

**Acknowledgements.** The authors would like to thank Prof. Alexander Pukhov for the use of the PIC code VLPL. This work was supported by the Strategic Priority Research Program of the Chinese Academy of Sciences (Grant No. XDB 16010000) and the National Natural Science Foundation of China (Grants No. 11875307 and No.11935008).

**Disclosures.** The authors declare no conflicts of interest.